\documentclass[aps,prl, twocolumn, 10pt,superscriptaddress]{revtex4-2}
\usepackage[dvipsnames]{xcolor}
\usepackage{mathrsfs}
\usepackage{natbib,hyperref}
\setcitestyle{square,sort&compress,comma,numbers}
\hypersetup{colorlinks=true, citecolor=blue, urlcolor=blue, linkcolor=blue}
\usepackage[english]{babel}
\usepackage[utf8]{inputenc}
\usepackage{graphicx}
\usepackage[caption=false]{subfig}
\usepackage{amsmath}
\usepackage{amsfonts}
\usepackage{amssymb}
\usepackage{color,soul}
\setulcolor{red}
\usepackage{easyReview}
\usepackage{xcolor}

\usepackage[normalem]{ulem}
\begin{document}

\title{Hydrodynamic Attraction and Hindered Diffusion Govern First-passage Times of Swimming Microorganisms}
\author{Yanis Baouche}
\affiliation{Max-Planck-Institut f{\"u}r Physik komplexer Systeme, N{\"o}thnitzer Stra{\ss}e 38,
01187 Dresden, Germany}
\author{Magali Le Goff}
\affiliation{Institut fur Theoretische Physik, Universit{\"a}t Innsbruck, Technikerstra{\ss}e 21A, A-6020 Innsbruck, Austria}
\author{Thomas Franosch}
\affiliation{Institut fur Theoretische Physik, Universit{\"a}t Innsbruck, Technikerstra{\ss}e 21A, A-6020 Innsbruck, Austria}
\author{Christina Kurzthaler}
\email{ckurzthaler@pks.mpg.de}
\affiliation{Max-Planck-Institut f{\"u}r Physik komplexer Systeme, N{\"o}thnitzer Stra{\ss}e 38,
01187 Dresden, Germany}
\affiliation{Center for Systems Biology Dresden, Pfotenhauerstra{\ss}e 108, 01307 Dresden, Germany}
\affiliation{Cluster of Excellence, Physics of Life, TU Dresden, Arnoldstra{\ss}e 18, 01062 Dresden, Germany}

\begin{abstract} 
The motion of microorganisms in their natural habitat is strongly influenced by their propulsion mechanisms, geometrical constraints, and random fluctuations. Here, we study numerically the first-passage-time (FPT) statistics of microswimmers, modeled as force-dipoles, to reach a no-slip wall. Our results demonstrate that hindered diffusion near the wall can increase the median FPT by orders of magnitude compared to ``dry'' agents, while the intricate interplay of active motion and hydrodynamic attraction speeds up the arrival at large  P{\'e}clet numbers (measuring the importance of self-propulsion relative to diffusion). Strikingly, it leads to a non-monotonic behavior as a function of the dipole strength, where pushers reach the wall significantly faster than pullers. The latter become slower at an intermediate dipole strength and are more sensitive to their initial orientation, displaying a highly anisotropic behavior. 
\end{abstract} 

\maketitle

Microorganisms have evolved ingenious strategies for surviving and performing a multitude of biological functions, ranging from bacterial infections~\cite{ocroininHostEpithelialCell2012, otteStatisticsPathogenicBacteria2021a} to fertilization~\cite{suarezSpermTransportFemale2006b, zaferaniBiphasicChemokinesisMammalian2023}. Many reside in aqueous settings and employ swimming mechanisms to navigate their surroundings~\cite{laugaHydrodynamicsSwimmingMicroorganisms2009a, bechingerActiveParticlesComplex2016a, laugaFluidDynamicsCell2020a, kurzthalerOutofequilibriumSoftMatter2023a}, where the interplay of viscous forces and strong stochastic fluctuations dictates their dynamics. These processes are expected to play a fundamental role for their ability to find specific targets and their approach towards boundaries~\cite{berkeHydrodynamicAttractionSwimming2008a}, where they accumulate to withstand harsh conditions~\cite{hannigOralCavityKey2009, drescherBiofilmStreamersCause2013a, wongetal.RoadmapEmergingConcepts2021, kurzMorphogenesisBiofilmsPorous2023}. Long-ranged hydrodynamic interactions with the target can generate attraction towards its boundary and reorientation~\cite{spagnolieHydrodynamicsSelfpropulsionBoundary2012e, spagnolieGeometricCaptureEscape2015a}, while modifying the noise strength~\cite{bian111YearsBrownian2016b}. These aspects render the study of the first-passage-time (FPT) statistics, capturing how long these agents require to find their target, a fascinating topic of research with far reaching implications for both biology and nanotechnology, which aims at designing efficient cargo carriers~\cite{erkocMobileMicrorobotsActive2019a, alapanMicroroboticsMicroorganismsBiohybrid2019a, volpeRoadmapAnimateMatter2024}. 

Theoretical insights come from two main approaches, where the first emanates from entailing the active agents with some internal ``intelligence''~\cite{piroOptimalNavigationStrategies2021, piroEfficiencyNavigationStrategies2022, piroOptimalNavigationMicroswimmers2022a, daddi-moussa-iderHydrodynamicsCanDetermine2021a, hartlMicroswimmersLearningChemotaxis2021, colabreseFlowNavigationSmart2017, zouGaitSwitchingTargeted2022} and the second deals with FPT statistics of different stochastic processes describing microswimmer motion at the mesoscale~\cite{viswanathanPhysicsForagingIntroduction2011a, weissApplicationsPersistentRandom2002, angelaniFirstpassageTimeRunandtumble2014, malakarSteadyStateRelaxation2018a, gueneauRunandtumbleParticleOnedimensional2025a}. In the first approach, using control theory optimal strategies for reaching a target have been identified in different force fields~\cite{piroOptimalNavigationStrategies2021, piroEfficiencyNavigationStrategies2022, piroOptimalNavigationMicroswimmers2022a} and near boundaries~\cite{daddi-moussa-iderHydrodynamicsCanDetermine2021a}, where hydrodynamic interactions can crucially impact the fastest path towards a target. Furthermore, recent machine learning approaches, relying on genetic algorithms~\cite{hartlMicroswimmersLearningChemotaxis2021} and reinforcement learning~\cite{colabreseFlowNavigationSmart2017,zouGaitSwitchingTargeted2022}, have been developed to endow active particles with intrinsic mechanisms for their gait control towards target locations. 

Standard target-search problems, on the other hand, have been put forward in the 20th century in order to quantify chemical kinetics~\cite{gudowska-nowakPrefaceMarianSmoluchowskis2017,grebenkovTargetSearchProblems2024a}, and rely on deriving FPT statistics for specific dynamics.  In the context of ``dry'' active agents, where hydrodynamics are ignored, these properties have been addressed analytically for run-and-tumble particles in one dimension~\cite{weissApplicationsPersistentRandom2002, angelaniFirstpassageTimeRunandtumble2014, malakarSteadyStateRelaxation2018a, gueneauRunandtumbleParticleOnedimensional2025a}. Additionally, they have been studied analytically~\cite{ditrapaniActiveBrownianParticles2023c,baoucheFirstpassagetimeStatisticsActive2025} and numerically~\cite{basuActiveBrownianMotion2018a} for active Brownian particles hitting a wall in two dimensions, showing anisotropic behavior with respect to the initial swimming direction. Yet, the aspect of hydrodynamic interactions has, despite its abundance in many biological systems, not been accounted for. 

While the detention times of noisy microswimmers near walls depend strongly on the details of the hydrodynamic interactions~\cite{schaarDetentionTimesMicroswimmers2015a}, the impact of noise in such situations remains largely unexplored. Hydrodynamic studies, ignoring the aspect of stochastic fluctuations, have demonstrated intriguing physics, such as accumulation~\cite{berkeHydrodynamicAttractionSwimming2008a, elgetiWallAccumulationSelfpropelled2013, spagnolieGeometricCaptureEscape2015a, prakashFeedersExpellersTwo2024} and circular near-surface motion~\cite{laugaSwimmingCirclesMotion2006a,elgetiHydrodynamicsSpermCells2010}, emerging from the coupling of the velocity field produced by the microswimmers with boundaries. Thus, establishing a fundamental understanding of both hydrodynamics and noise will be of paramount importance for revealing insights into different aspects of microbiology, such as the transition from a planktonic swimming state to a surface-attached biofilm and target-search-efficiency at the microscale, relevant for advancing biomedical applications. 

Here, we study the FPT statistics of swimming microorganisms 
by accounting for the hydrodynamic interactions through both the flow field produced by the agents and the spatially-varying diffusivities in terms of a far-field description. Our results reveal a drastic slowdown due to hydrodynamically-hindered diffusion close to the wall, which can be compensated by the hydrodynamic attraction of pusher-like microswimmers that become faster than their ``dry" counterpart at high P{\'e}clet numbers. Flow fields produced by pullers lead to even longer FPTs, due to their enhanced hydrodynamic repulsion. We thus demonstrate how the swimming gait and slowdown of diffusion determine the efficiency of active target-search processes.

\begin{figure}[tp]
\includegraphics[width=0.5\textwidth]{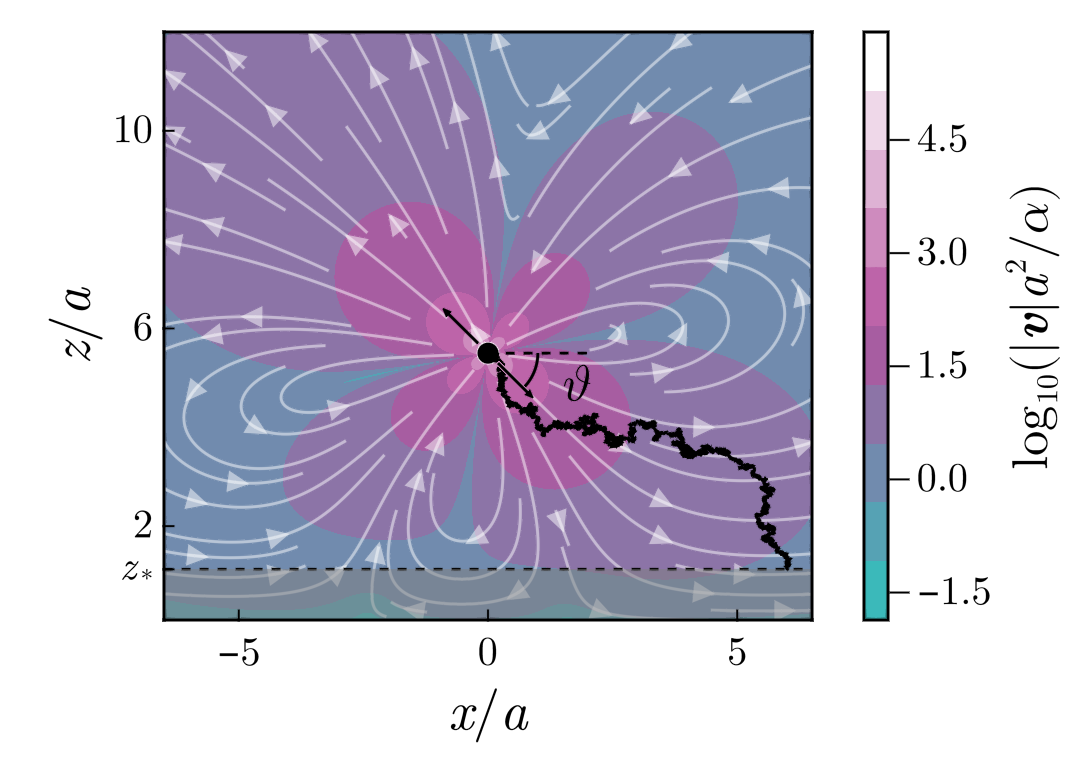}
\caption{Velocity field $\boldsymbol{v}$ induced by the swimmer with force dipole strength~$\hat{\alpha} = 4$ and orientation $\vartheta=-\pi/4$, shown in the  $(xOz)$ plane. The black line is an exemplary trajectory. The heatmap represents the magnitude of the velocity field. Further,  $z^{*}$ represents the target height.  \label{fig:flow_field}}
\end{figure}

\paragraph{Model.--} We consider an active (spherical) particle swimming near a wall in a quasi-steady low-Reynolds number flow in three dimensions. The swimmer has a radius $a$ and is oriented along the swimming direction~$\boldsymbol{e}$. 
We describe the flow field produced by the swimmer at distances $z\gg a$ in terms of a force-dipole along~$\boldsymbol{e}$ of strength~$\alpha$. The spatially-varying velocity and pressure fields, $\boldsymbol{v}(\boldsymbol{r})$ and $p(\boldsymbol{r})$, obey the Stokes equations: $\mu\nabla^2\boldsymbol{v}= \boldsymbol{\nabla}p$  and $\boldsymbol{\nabla}\cdot\boldsymbol{v}= 0$, where $\mu$ denotes the fluid dynamic viscosity.
Using the image method, the velocity fields have been obtained analytically~\cite{blakeFundamentalSingularitiesViscous1974a, spagnolieHydrodynamicsSelfpropulsionBoundary2012e}: $\boldsymbol{v}= \boldsymbol{v}^{u}+\boldsymbol{v}^*$, where $\boldsymbol{v}^{u}$ corresponds to the velocity field produced by a force dipole in an unbounded domain and $\boldsymbol{v}^*$ is the (image) contribution due to the nearby wall (Fig.~\ref{fig:flow_field}). The latter affects the instantaneous position~$\boldsymbol{r}(t)$ and orientation~$\boldsymbol{e}(t)$ of the active agent via 
\begin{subequations}
\begin{align}
    \frac{\mathrm{d} \boldsymbol{r}}{\mathrm{d} t} &= v\boldsymbol{e} + \boldsymbol{v}^\mathrm{HI} + \boldsymbol{v}^\mathrm{noise}+\boldsymbol{v}^{\mathrm{drift}}, \label{eq:dynamics_position}\\
    \frac{\mathrm{d} \boldsymbol{e}}{\mathrm{d} t} &= \left(\boldsymbol{\omega}^\mathrm{HI}+\boldsymbol{\omega}^\mathrm{noise}+\boldsymbol{\omega}^{\mathrm{drift}}\right)\wedge\boldsymbol{e}\label{eq:dynamics_orientation},
\end{align}
\end{subequations}
which contain contributions from the hydrodynamic interactions with the wall ($\boldsymbol{v}^\mathrm{HI}$ and $\boldsymbol{\omega}^\mathrm{HI}$) and Brownian noise ($\boldsymbol{v}^\mathrm{noise}$ and $\boldsymbol{\omega}^\mathrm{noise}$), and their associated  drifts ($\boldsymbol{v}^\mathrm{drift}$ and $\boldsymbol{\omega}^\mathrm{drift}$). The hydrodynamic contributions for the translational and angular velocities are obtained via Fax{\'e}n's law~\cite{lealAdvancedTransportPhenomena2007}
\begin{align}
\boldsymbol{v}^\mathrm{HI} &= \boldsymbol{v}^*(\boldsymbol{r}) \quad \mathrm{and} \quad \boldsymbol{\omega}^\mathrm{HI} = \frac{1}{2}\boldsymbol{\nabla}\wedge \boldsymbol{v}^*(\boldsymbol{r}). 
\end{align}
Assuming that the swimmer is described by a force dipole, we have, in the cartesian basis $(\boldsymbol{\hat{x}},\boldsymbol{\hat{y}},\boldsymbol{\hat{z}} )$ 
\begin{subequations}
\begin{align}
\boldsymbol{v}^\mathrm{HI} &= \frac{3\alpha}{16z^2}\left(1-3\cos(2\vartheta)\right)\boldsymbol{\hat{z}}+\frac{3\alpha}{8z^2}\sin(2\vartheta)\boldsymbol{\hat{\phi}},\\ 
\boldsymbol{\omega}^\mathrm{HI} &= \frac{3\alpha}{16z^3}\sin(2\vartheta)\boldsymbol{\hat{\phi}}^{\perp} ,
\end{align}
\end{subequations}
with pitch angle $\vartheta$, measuring the angle relative to the nearby wall, $\boldsymbol{\hat{\phi}}= \cos(\phi)\boldsymbol{\hat{x}} + \sin(\phi)\boldsymbol{\hat{y}}$, and $\boldsymbol{\hat{\phi}}^{\perp} = \boldsymbol{\hat{z}} \wedge \boldsymbol{\hat{\phi}}$. 

\begin{figure*}[htp]
\includegraphics[width=1.0\textwidth]{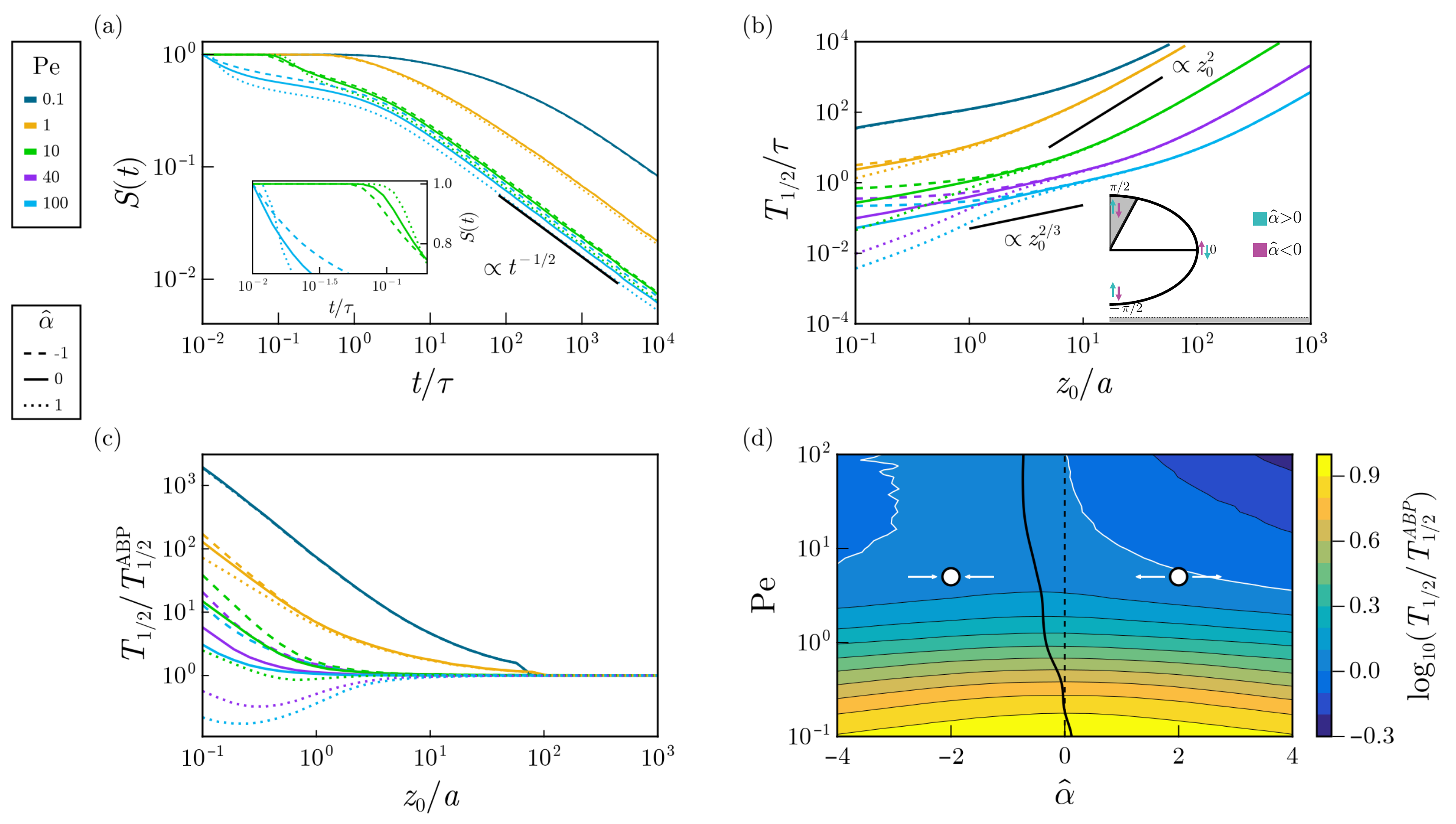}
\caption{\label{fig:2_ratio_pp} (a) Surival probability $S(t)$ as a function of time $t$ for several P{\'e}clet numbers $\mathrm{Pe}$ and reduced dipole strengths~$\hat{\alpha}$. Here, the initial distance is $z_{0}/a=2.0$. (\textit{Inset}) shows a magnification of $S(t)$ at short times. (b) Median FPT $T_{1/2}$ as a function of the initial distance $z_{0}$. (\textit{Inset}) Schematic of how the hydrodynamic-induced velocities affect the swimmer. Arrows pointing downwards indicate
attraction while arrows pointing upwards indicate repulsion. The gray-shaded area represents orientations of agents departing from the wall and not reaching it.  (c) Ratio $T_{1/2}/T_{1/2}^{\mathrm{ABP}}$ of the median FPT as a function of the initial distance $z_{0}$ for several P{\'e}clet numbers $\mathrm{Pe}$ and reduced dipole strengths $\hat{\alpha}$. (d) Contour plot of the ratio $T_{1/2}/T_{1/2}^{ABP}$ as a function of $\mathrm{Pe}$ and $\hat{\alpha}$ for $z_{0}/a$=5. White lines indicate $T_{1/2}= T_{1/2}^{ABP}$ and the black line corresponds to the maximum $\mathrm{max} \ [T_{1/2}/T_{1/2}^{ABP}]$ for different $\mathrm{Pe}$-numbers.}
\end{figure*}

Next, we introduce the contributions due to Brownian motion, assuming that the diffusivities are the same as for their passive counterparts:
\begin{align}
\boldsymbol{D}=k_BT\boldsymbol{M} &= k_B T\begin{pmatrix} \boldsymbol{M}^{t} & \boldsymbol{M}^{tr}\\
\boldsymbol{M}^{rt} & \boldsymbol{M}^{r}
\end{pmatrix}, 
\end{align}
where $k_{B}$ denotes the Boltzmann constant, $T$ the temperature, and the matrices $\boldsymbol{M}^{t, r, tr, rt}$~\cite{e.uspalBoundaryElementMethod2019} represent the mobilities for translation, rotation, and the translational/rotational coupling.  In the far-field regime ($a/z\ll1$), to leading order the components of the diffusion tensor are 
\begin{subequations}
\begin{align}
    D_{\perp}^t(z) &= D_0^t \left [1 -  \frac{9}{8} \left (\frac{a}{z} \right )  \right ], \\
    D_{\parallel}^t(z) &= D_0^t \left [1 -  \frac{9}{16} \left (\frac{a}{z} \right )   \right ],  \\
    D_{\perp}^r(z) &= D_0^r \left [1 - \frac{1}{8} \left (\frac{a}{z} \right )^3  \right ], \\
    D_{\parallel}^r(z) &= D_0^r \left [1 - \frac{5}{16} \left (\frac{a}{z} \right )^3   \right ],
\end{align}
\end{subequations}
where $\perp$ and $\parallel$ correspond to the direction perpendicular and parallel to the wall, respectively. The coefficients $D_{0}^{t}$ and $D_{0}^{r}$ are the diffusion coefficients in bulk. In a first approximation, we neglect the translational/rotational coupling  $\boldsymbol{D}^{rt}= \boldsymbol{D}^{tr}=\boldsymbol{0}$ and focus on the wall-induced effect on translational and rotational diffusion, such that the diffusion matrix (for a sphere) is diagonal 
\begin{equation}
     \boldsymbol{D}= \mathrm{diag} \left(D^{t}_{\parallel}(z),D^{t}_{\parallel}(z),D_{\perp}^{t}(z), D^{r}_{\parallel}(z),D^{r}_{\parallel}(z),D^{r}_{\perp}(z) \right),
\end{equation}
and the Brownian noise is given by: 
\begin{align}
(\boldsymbol{v}^{\mathrm{noise}}, \boldsymbol{\omega}^{\mathrm{noise}})^T &= \sqrt{2\boldsymbol{D}}\boldsymbol{\xi},
\end{align}
where $\boldsymbol{\xi}$ is Gaussian white noise of zero mean and delta-correlated variance, $\langle \xi_i(t)\xi_j(t')\rangle = \delta_{ij}\delta(t-t')$ for $i,j=1,...,6$. Near a planar wall, the diffusive processes are anisotropic and $z$-dependent ($\boldsymbol{D} = \boldsymbol{D}(z)$) and thus drift terms emerge [see Supplementary Information (SI)~\footnote{See Supplemental Material at XX, which includes Refs.~\cite{lealAdvancedTransportPhenomena2007, spagnolieHydrodynamicsSelfpropulsionBoundary2012e, chwangHydromechanicsLowReynoldsnumberFlow1975, kurzthalerIntermediateScatteringFunction2016, happelLowReynoldsNumber1991, jonesRotationalDiffusionTracer1992a}, for details on the theory and simulations.}]: 
    \begin{align}
        \boldsymbol{v}^{\mathrm{drift}} &= D_{0}^{t}\frac{9a}{8z^{2}} \hat{\boldsymbol{z}}  \label{eq:drift_position_dim} \quad \text{and}\quad 
        \boldsymbol{\omega}^{\mathrm{drift}} = \boldsymbol{e} \wedge \boldsymbol{D}^{r} \cdot \boldsymbol{e}.
    \end{align}
The dynamics of the agent is characterized by two nondimensional numbers: the P{\'e}clet number $\mathrm{Pe}=va/D_{0}^{t}$, characterizing the relative importance
of active motion and diffusion, as well as the reduced dipole strength $\hat{\alpha}= \alpha/(va^{2})$. We use the hydrodynamic radius $a=\sqrt{3D_0^t/(4D_0^r)}$ as length scale. Here, we consider pushers $\hat{\alpha}>0$, having a flow-signature as in Fig.~\ref{fig:flow_field}, pullers $\hat{\alpha}<0$, and neutral swimmers $\hat{\alpha}=0$, whose hydrodynamic interaction with the wall only appears in the diffusivities. We compute the FPT properties of agents starting at a distance $z(0)=z^{*}+z_0$ from the boundary, where $z^{*}$ denotes the target distance. Unless stated otherwise, agents are oriented at random initial pitch angles~$\vartheta_{0}$. 

\paragraph{Hydrodynamic attraction speeds up pushers. --} The FPT statistics are typically studied through the survival probability $S(t)$ (i.e., the probability that the agent has not reached the target distance $z^{*}$ up to time $t$). In an unconfined domain, it commonly behaves as $S(t) \sim t^{-\sigma}$ at long times, with exponent $\sigma$ depending on the underlying dynamics~\cite{levernierSurvivalProbabilityStochastic2019}. Despite hydrodynamic interactions, our simulations [Fig.~\ref{fig:2_ratio_pp}(a)] show $\sigma=1/2$, just like the ``dry'' counterpart with small activity~\cite{baoucheFirstpassagetimeStatisticsActive2025}. However, the short-time behavior and the tail's amplitude depend sensitively on the P{\'e}clet number. On the one hand, a diffusion-dominated dynamics $(\mathrm{Pe}=0.1)$ will lead to a lower probability to be absorbed at short-times because diffusion is hindered close to the boundary. On the other hand, the regime of high activity ($\mathrm{Pe} \gg 1$) displays a higher absorption probability, as persistent motion becomes the main locomotary mechanism, thus allowing the agent to circumvent slow diffusion. 
We further note that the impact of the flow signature ($\hat{\alpha}$) becomes mainly apparent at short times and will become clear later. Additionally, as a consequence of the slow decay of the survival probability, the mean FPT (MFPT) diverges, therefore not giving any information about the completion time. Consequently, we use the median $T_{1/2}$ as a metric for the FPT, which has the advantage of almost always being defined, being immune to fluctuations and outliers, and most importantly to provide a quantitative description of a process' speed of completion~\cite{belanMedianModeFirst2020a}.

We first consider the median $T_{1/2}$ for a range of $\mathrm{Pe}$ and $\hat{\alpha}$ as a function of initial distance $z_0/a$ to study how the interplay of all processes affects the absolute time it takes the agent to reach the boundary [Fig.~\ref{fig:2_ratio_pp} (b)]. 
For neutral swimmers ($\hat{\alpha}=0$) the median FPT scales as~$z_{0}^{2/3}$ at distances smaller than the persistence length $z_0\lesssim \ell_{p}=v/D_{0}^{r}$. This behavior remains for constant diffusivities [Fig.2 in SI] and can thus be attributed to the coupling between persistent motion and diffusive rotational motion. Indeed, an ABP typically displays a superballistic regime in its mean-squared displacement $\langle \Delta z(t)^{2} \rangle \propto t^{3}$ through the influence of its initial angle~\cite{tenhagenBrownianMotionSelfpropelled2011b}. The latter is captured by the median, as it does not pick up particles that do not reorient to eventually reach the wall. Thus, the typical time to reach the boundary from a distance $z_0$ leads to a median such that $T_{1/2} \propto z_{0}^{2/3}$. At large initial distances ($z_{0} \gg \ell_p$), the median behaves as $T_{1/2} \propto z_{0}^{2}$, which reflects the long-time effective diffusive regime of an active Brownian particle. Importantly, in this regime the dipole strength has no impact and all medians associated with a certain $\mathrm{Pe}$ coalesce into a single curve.

At short and intermediate initial distances $z_{0}\lesssim \ell_{p}$, the hydrodynamic contributions allow differentiating the type of swimmer [Fig.~\ref{fig:2_ratio_pp} (b)]. In particular, the prominent feature that emerges is that pushers almost always exhibit a smaller median than pullers when the initial angle is randomized. Inspecting the survival probability,  our results show that pullers reach the wall in two populations: the first swimming towards the wall and actually reaching it faster than pushers, which becomes manifest in a smaller survival probability [Fig.~\ref{fig:2_ratio_pp}(a) \textit{inset}]; the second population departs away from the wall and reaches it after a very long time due reorientation towards the wall via rotational diffusion (or never), leading to a slower decay of $S$ at long times. The latter scenario eventuates in spite of the hydrodynamic attraction for pullers, as persistent motion $v  \gtrsim v^{\mathrm{HI}}_z$ (corresp. to $1\gtrsim3|\alpha|/(4vz^{2})$) prevails. 

To understand the differences between pushers and pullers we elucidate the distinct contributions due to the hydrodynamic coupling. First, neglecting the hydrodynamic reorientation by setting $\boldsymbol{\omega}^{\mathrm{HI}}=\boldsymbol{0}$ does not change the results, while for $\boldsymbol{v}^{\mathrm{HI}}=\boldsymbol{0}$ the results for various $\hat{\alpha}$ are almost indistinguishable (see Fig.6 in SI), indicating that hydrodynamic attraction is what causes the difference in the medians. We argue that this discrepancy results from the fact that out of half of the population that reach the wall at long times $t\gtrsim T_{1/2}$, most agents first swim persistently away from the wall and thus an unbalance of hydrodynamic attraction arises between pushers and pullers. In particular, pullers from a certain part of their attractive region are not represented in the median (corresponding to the gray-shaded area in the inset of Fig.~\ref{fig:2_ratio_pp} (b)). If we let $\vartheta^{*} \in [0,\pi/2]$, then averaging the attraction over the orientations $\int_{-\pi/2}^{\vartheta^\star} V_{z}(\vartheta) \sin(\vartheta)~\mathrm{d}\vartheta \propto - \alpha/z^{2}$ shows that pushers experience a net attraction and pullers a net repulsion. Overall, this demonstrates that hydrodynamic attraction is the prime cause of the speed up of pushers, while their reorientation plays a minor role. It is important to note that hydrodynamic reorientation becomes important for post-arrival phenomena~\cite{laugaSwimmingCirclesMotion2006a, spagnolieHydrodynamicsSelfpropulsionBoundary2012e} and dictate their detention times~\cite{schaarDetentionTimesMicroswimmers2015a}.


\textit{Slowdown due to hindered diffusion.--} In order to obtain a quantitative idea of how hydrodynamic interactions and height-dependent diffusion affect the median, we compare $T_{1/2}$ with that of a simple ABP~$T_{1/2}^{\mathrm{ABP}}$ ($\hat{\alpha}=0$ and constant bulk diffusivities~$D_{0}^{t}, D_{0}^{r}$). Figure~\ref{fig:2_ratio_pp}(c) shows an increase by up to two orders of magnitude of the median FPT -- a significant slowdown in comparison to ``dry" agents -- revealing that hindered diffusion at the boundary is the leading cause of this slowdown, even at high P{\'e}clet numbers ($\mathrm{Pe} =100$). We further note that we have also split the contributions of translational and rotational diffusion and observed that the latter does not influence the FPT statistics (see Sec.~II.B in SI), thus supporting our first approximation, in which the rotation-translation coupling  term (of same order) was neglected. 

\textit{Non-monotonic behavior.--}To unravel the interplay between activity and flow signature, we further plot the same ratio as before as a function of $\mathrm{Pe}$ and $\hat{\alpha}$ for $z_{0}/a=5$ [Fig.~\ref{fig:2_ratio_pp} (d)]. Through the black line representing the maximum of $T_{1/2}/T_{1/2}^{\mathrm{ABP}}$, we observe that pullers are slightly faster than pushers at very low P{\'e}clet number ($\mathrm{Pe}\ll 1$), as activity is too low to prevent the hydrodynamic attraction at these distances. Most prominently, we find that this picture reverses for increasing activity and pushers become faster than pullers.  Moreover, larger dipole strength $\hat{\alpha}$ can decrease the median FPT due to hydrodynamic attraction of both pullers and pushers, leading to a non-monotonic behavior of the median with a maximum at $\hat{\alpha}\lesssim 0$. This indicates that at an intermediate dipole strength, pullers take longer to reach the wall: for large $\hat{\alpha}$ hydrodynamic attraction takes over active motion away from the wall ($v \lesssim v^{\mathrm{HI}}_z$ (corresp. to $1\lesssim 3|\alpha|/(4vz^{2})$)), while for small $\hat{\alpha}$ the behavior of a neutral swimmer is approached.  

\begin{figure}[tp]
\includegraphics[width=0.5\textwidth]{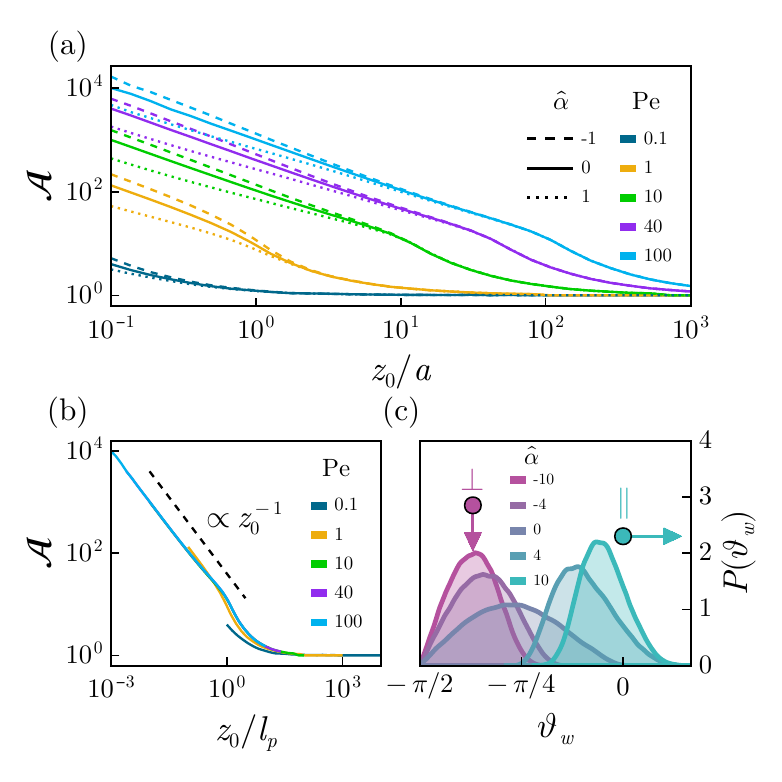}
\caption{\label{fig:3_aniso}(a) Anisotropy $\mathcal{A}$ as a function of the initial distance~$z_{0}$ for several P{\'e}clet numbers $\mathrm{Pe}$ and dipole strengths~$\hat{\alpha}$. (b)~Anisotropy rescaled by the persistence length $l_{p}$ for $\hat{\alpha}=0$. (c) Distribution of the arrival angles $\vartheta_{w}$ as a function of the dipole strength $\hat{\alpha}$. Here, $z_{0}/a=5.6$ and $\mathrm{Pe}=10$.}
\end{figure}

\textit{Impact of initial orientation.--} Our results suggest that the initial orientation plays a major role in the FPT statistics, similar to the dry counterpart~\cite{baoucheFirstpassagetimeStatisticsActive2025}. To quantify its role, we introduce an anisotropy metric 
\begin{align}
    \mathcal{A} = \frac{T_{1/2}(z_{0}, \pi/2)}{T_{1/2}(z_{0}, -\pi/2)},
\end{align}
measuring the ratio of the median FPTs of swimmers departing from and swimming towards the wall. Figure~\ref{fig:3_aniso}(a) shows that the anisotropy decays with the initial distance for all activities and dipole strengths, and relaxes exponentially to zero for $z_{0}/a \gg 1$, where the memory of the initial angle is progressively lost due to rotational diffusion. We note that for $\hat{\alpha}=0$ rescaling the initial distance by the persistence length  collapses our data to a single curve where $\mathcal{A}\propto z_0^{-1}$ up to $z_{0}/l_{p} \simeq 1$. This is due to agents facing the wall and reaching it ballistically ($\propto z_{0}$), while those departing away from it exhibit a median that plateaus until $z_{0} \simeq l_{p}$ (see Sec. IIC in SI). Their ratio thus linearly decays with initial distance. At $z_{0}/l_{p} \gtrsim 1$ rotational diffusion typically decorrelates orientations as $\propto e^{-2D_{0}^{r}t}$ and the anisotropy vanishes.

Additionally, we note that pullers ($\hat{\alpha}=-1$) display a bigger anisotropy than pushers ($\hat{\alpha}=1$). This can be once again explained by the fact that pullers swimming towards the wall experience attraction, making them faster than pushers in this instance. In the departing scenario, pushers experience boundary-aligning torques, making them slightly faster, such that the overall anisotropy is more pronounced for pullers. The anisotropic behavior can be further understood by studying the angle at which they reach the boundary $\vartheta_{w}$. Figure~\ref{fig:3_aniso}(c) shows that while pushers ($\hat{\alpha} >0$) tend to approach the wall in a parallel manner, pullers ($\hat{\alpha} <0$) swim straight towards the wall due to hydrodynamic reorientation and subsequent attraction, in agreement with the orientational dynamics of force-dipole swimmers near walls~\cite{berkeHydrodynamicAttractionSwimming2008a,spagnolieHydrodynamicsSelfpropulsionBoundary2012e,goldsteinGreenAlgaeModel2015}.

\paragraph{Summary and Conclusions.--}
Through the study of the FPT properties of different types of swimming organisms, we demonstrated that the time it takes an active agent to reach a boundary is dictated by the interplay of their swimming gait and diffusion. Using numerical simulations, we find that although hindered diffusion always obstructs wall absorption, hydrodynamic interactions can make the swimmer faster or slower than a ``dry" agent, with the outcome dependent on its flow signature. 
In particular, persistent swimming disrupts the overall balance of wall-induced attraction and repulsion, resulting in a net decrease of the median FPT for pushers and increase for pullers. This behavior is magnified closer to the wall, where diffusion becomes negligible. Our results suggest that for an {\it Escherichia coli} bacterium, having a reduced dipole strength of $\hat{\alpha}\approx 1.5$ and a P{\'e}clet number of $\mathrm{Pe}= O(100)$ \cite{drescherFluidDynamicsNoise2011,kurzthalerCharacterizationControlRunandTumble2024b}, hydrodynamic interactions could expedite their approach towards walls, which remains to be tested in experiments. 

Moreover, the angle at which the microorganism initially starts to swim is key in finding the target. Quantifying this directional bias through an anisotropy measure, we observe that while agents originally pointing towards the wall naturally reach it faster, this effect solely occurs at distances smaller than their persistence length, at which point the memory of the initial orientation starts fading away. In this context, it also plays a larger role for pullers, as attraction and persistent swimming play in concert to make them hit the wall faster than pushers  when facing towards it.

In this work, we used a far-field description for the swimmer, but a theory that resolves the microscopic details of the swimmer would be required to resolve its behavior close to the boundary, as locomotory mechanisms such as the beating of flagella or cilia would be a deciding element at such small length scales~\cite{martinColiBacteriumTumbling2025}. However, analytically resolving the near-field through an asymptotic matching with lubrication theory or using bispherical coordinates is only expected to yield interface effects such as scattering and oscillations~\cite{crowdyTreadmillingSwimmersNoslip2011, ishimotoDynamicsTreadmillingMicroswimmer2017}, but we do not expect qualitative changes for the FPT statistics.

Our work sheds light on how confined hydrodynamics affects the approach of active agents towards boundaries and is expected to provide grist for future microbiological target-search studies. We anticipate that it will allow for a broader consideration of the role of hydrodynamics in first-passage problems for active particles, which are often only studied through the lens of statistical physics due to the difficulty in resolving hydrodynamic interactions. For biological swimmers, arrival at a boundary represents the first step towards the emergence of important effects such as biofilm formation~\cite{sharmaMicrobialBiofilmReview2023,zhaoUnderstandingBacterialBiofilms2023}, material degradation~\cite{dattaMicrobialInteractionsLead2016, slomkaEncounterRatesBacteria2020,alcolombriSinkingEnhancesDegradation2021}, microbial predation~\cite{saleemDiversityProtistsBacteria2013,tusonBacteriaSurfaceInteractions2013} or symbiosis~\cite{weiland-brauerFriendsFoesMicrobial2021}, and we thus expect our study to contribute to their understanding.

\paragraph{Acknowledgments.--}
Y.B. gratefully acknowledges discussions with Sagnik Garai and Akhil Varma. This research was funded in part by the Austrian Science Fund (FWF) 10.55776/P35580.

\bibliography{bibliography_main}

\newpage

\end{document}


\title{Supplemental Material: Hydrodynamic Attraction and Hindered Diffusion Govern First-passage Times of Swimming Microorganisms}
\author{Yanis Baouche}
\affiliation{Max-Planck-Institut f{\"u}r Physik komplexer Systeme, N{\"o}thnitzer Stra{\ss}e 38,
01187 Dresden, Germany}
\author{Magali Le Goff}
\affiliation{Institut fur Theoretische Physik, Universit{\"a}t Innsbruck, Technikerstra{\ss}e 21A, A-6020 Innsbruck, Austria}
\author{Thomas Franosch}
\affiliation{Institut fur Theoretische Physik, Universit{\"a}t Innsbruck, Technikerstra{\ss}e 21A, A-6020 Innsbruck, Austria}
\author{Christina Kurzthaler}
\email{ckurzthaler@pks.mpg.de}
\affiliation{Max-Planck-Institut f{\"u}r Physik komplexer Systeme, N{\"o}thnitzer Stra{\ss}e 38,
01187 Dresden, Germany}
\affiliation{Center for Systems Biology Dresden, Pfotenhauerstra{\ss}e 108, 01307 Dresden, Germany}
\affiliation{Cluster of Excellence, Physics of Life, TU Dresden, Arnoldstra{\ss}e 18, 01062 Dresden, Germany}

\maketitle

\tableofcontents

\begin{figure}[h]
    \centering
    \includegraphics[width=0.7\linewidth]{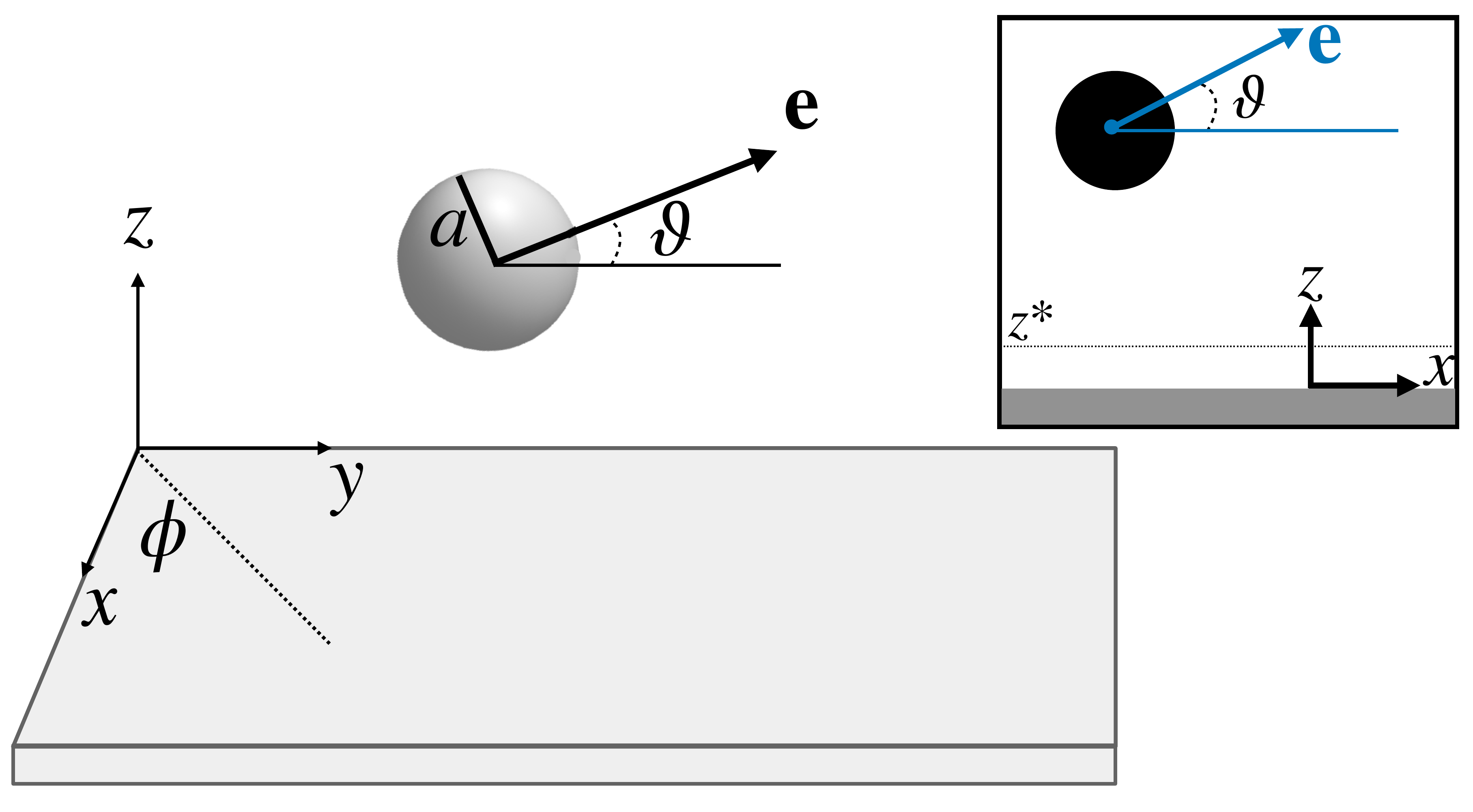}
    \caption{Model set-up of a spherical swimmer near a wall. The coordinate system is defined to be $\{O,x,y,z\}$, where $z^{*}=1.125$ corresponds to the target position (see inset). The agent has radius $a$ and swims along its orientation $\boldsymbol{e}$ parametrized by the pitch angle $\vartheta$ and the azimuthal angle $\phi$.}
    \label{fig:schematic}
\end{figure}

\section{Model details}

In this section, we present details of our model, including the derivation of the hydrodynamic contributions and the spurious drift terms arising for spatially-varying diffusivities. 

\subsection{Far-field hydrodynamics \label{app:ff_hydro}}
Here, we expand on the flow field produced by a force dipole and on the image solution. Therefore, we first introduce the spatially-varying pressure and velocity fields produced by a point force, $p_F(\boldsymbol{r})$ and $\boldsymbol{v}_F(\boldsymbol{r})$, respectively, and the fluid viscosity $\mu$. We obtain the corresponding Green’s function to the Stokes equation in free space by placing a point force at $\boldsymbol{r}_{0}=(x_{0},y_{0},z_{0})$ and solving
\begin{align}
    \boldsymbol{\nabla}p_F - \mu\boldsymbol{\nabla}^{2}\boldsymbol{v}_F    = \boldsymbol{F} \delta(\boldsymbol{r}-\boldsymbol{r}_{0}), \label{eq:stokes_point_force}
\end{align}
with $\boldsymbol{F}=F\boldsymbol{e}$ and $\boldsymbol{e}=(\cos(\vartheta) \cos(\phi),\cos(\vartheta) \sin(\phi),\sin(\vartheta))$. The solution for the velocity field with boundary condition $\boldsymbol{v}_F(|\boldsymbol{r}|\to \infty) = \boldsymbol{0}$ is given by~\cite{happelLowReynoldsNumber1991}
\begin{subequations}
    \begin{align}
        \boldsymbol{v}_F &= \frac{F}{8\pi \mu} \left( \frac{\boldsymbol{I}}{r-r_{0}} + \frac{(\boldsymbol{r}-\boldsymbol{r}_{0})(\boldsymbol{r}-\boldsymbol{r}_{0}) }{(r-r_{0})^{3}} \right) \cdot \boldsymbol{e} =: \frac{F}{8\pi \mu} \mathbb{G}_F(\boldsymbol{r},\boldsymbol{r}_{0}) \cdot \boldsymbol{e} =: \frac{F}{8\pi \mu} \boldsymbol{G}_F(\boldsymbol{r},\boldsymbol{r}_{0}; \boldsymbol{e}
        ),
    \end{align}
\end{subequations}
where $\boldsymbol{I}$ denotes the identity matrix. The solution for an axisymmetric force dipole is then obtained by placing two forces monopoles separated by a distance $l$ in the $\boldsymbol{e}$ direction, and pointing in the $\boldsymbol{e}$ and $-\boldsymbol{e}$ directions, respectively:
\begin{subequations}
    \begin{align}
        \boldsymbol{v} &= \frac{F}{8\pi \mu} \left[ \mathbb{G}_F\left(\boldsymbol{r}, \boldsymbol{r}_{0}+ \frac{l}{2}\boldsymbol{e}\right) \cdot \boldsymbol{e} + \mathbb{G}_F\left(\boldsymbol{r}, \boldsymbol{r}_{0}- \frac{l}{2}\boldsymbol{e}\right) \cdot (-\boldsymbol{e}) \right] \\
        &= \frac{F}{8\pi \mu} \left[ \boldsymbol{G}_F\left(\boldsymbol{r},\boldsymbol{r}_{0} + \frac{\ell}{2}\boldsymbol{e} ;\boldsymbol{e}\right) - \boldsymbol{G}_F\left(\boldsymbol{r}, \boldsymbol{r}_{0} - \frac{\ell}{2} \boldsymbol{e} ; \boldsymbol{e}\right) \right]\\
         &= \frac{F \ell}{8\pi\mu} \boldsymbol{e} \cdot \boldsymbol{\nabla}_{0} \boldsymbol{G}_F(\boldsymbol{r},\boldsymbol{r}_{0};\boldsymbol{e}) + O(l^{2}) \\ 
         &=: \alpha \boldsymbol{G}_{\mathrm{FD}}(\boldsymbol{r},\boldsymbol{r}_{0};\boldsymbol{e},\boldsymbol{e}) + O(l^{2}).
    \end{align}
\end{subequations}
The Green's function for a force dipole is thus obtained from the directional derivative of the point-force Green’s function with dipole strength $\alpha$ and reads
\begin{align}
    \boldsymbol{G}_{\mathrm{FD}}(\boldsymbol{\tilde{\boldsymbol{r}}}; \boldsymbol{e},\boldsymbol{e} )=  3 \frac{(\boldsymbol{e}\cdot \boldsymbol{\tilde{r}})^{2} \boldsymbol{\tilde{r}}}{\tilde{r}^{5}} - \frac{\boldsymbol{\tilde{r}}}{\tilde{r}^{3}},
\end{align}
where $\boldsymbol{\tilde{r}}=\boldsymbol{r}-\boldsymbol{r}_{0}$ and $\tilde{r}=|\boldsymbol{\tilde{r}}|$. To take into account the no-slip boundary, we resort to the method of images, effectively placing an image at position $\boldsymbol{r}_{0}^{*}=\boldsymbol{r}_{0}-2z_{0}\boldsymbol{\hat{z}}$ (at the opposite side of the wall). The solution of Eq.~\eqref{eq:stokes_point_force} with boundary condition $\boldsymbol{v}_F=\boldsymbol{0}$ on the wall reads
\begin{align}
    \boldsymbol{v}_F = \frac{F}{8\pi \mu} \Big( \boldsymbol{G}(\boldsymbol{\tilde{r}}; \boldsymbol{e}) + \boldsymbol{G}^{*}(\boldsymbol{\tilde{r}}^{*}; \boldsymbol{e}) \Big)
\end{align}
with the image Green's function for the force monopole~\cite{chwangHydromechanicsLowReynoldsnumberFlow1975, spagnolieHydrodynamicsSelfpropulsionBoundary2012e}
\begin{align}
    \boldsymbol{G}^{*}(\boldsymbol{\tilde{\boldsymbol{r}}}^{*}; \boldsymbol{e}) &=\cos(\vartheta) \left( -\boldsymbol{G}(\boldsymbol{\tilde{r}}^{*};\boldsymbol{\hat{\phi}}) + 2h \boldsymbol{G}_{\mathrm{FD}}(\boldsymbol{\tilde{\boldsymbol{r}}}^{*};\boldsymbol{\hat{\phi}}, \boldsymbol{\hat{z}}) -2h^{2}\boldsymbol{G}_{\mathrm{SD}}(\boldsymbol{\tilde{r}}^{*};\boldsymbol{\hat{\phi}})  \right) \notag \\
    &+ \sin(\vartheta) \left( -\boldsymbol{G}(\boldsymbol{\tilde{\boldsymbol{r}}}^{*}; \boldsymbol{\hat{z}}) - 2h \boldsymbol{G}_{\mathrm{FD}}(\boldsymbol{\tilde{\boldsymbol{r}}}^{*};\boldsymbol{\hat{z}}, \boldsymbol{\hat{z}}) +2h^{2}\boldsymbol{G}_{\mathrm{SD}}(\boldsymbol{\tilde{r}}^{*};\boldsymbol{\hat{z}})  \right),
\end{align}
where $\boldsymbol{\hat{\phi}}= \cos(\phi)\boldsymbol{\hat{x}} + \sin(\phi)\boldsymbol{\hat{y}}$ and  $\boldsymbol{\tilde{r}}^{*} = \boldsymbol{r} - \boldsymbol{r}_{0}^{*}$. Additionally,  we have introduced the Green's function for a source dipole, defined by
\begin{align}
    \boldsymbol{G}_{\mathrm{SD}}(\boldsymbol{\tilde{r}} ; \boldsymbol{e}) = 3\frac{\boldsymbol{\tilde{r}} (\boldsymbol{e}\cdot \boldsymbol{\tilde{r}} ) }{\tilde{r}^{5}}  - \frac{\boldsymbol{e}}{\tilde{r}^{3}}.
\end{align}
Obtaining the image Green's function for the force dipole  $\boldsymbol{G}_{\mathrm{FD}}^{*}$ is done by taking the derivative of the image solution of the point force. 
The full velocity field in the presence of the wall is then $\boldsymbol{v} = \alpha (\boldsymbol{G}_{\mathrm{FD}}+\boldsymbol{G}_{\mathrm{FD}}^{*})$.
Finally, Fax{\'e}n's law (see Eq.~(2) in the main text) is used to retrieve the zeroth-order contributions to the swimmer's translational and angular velocities~\cite{lealAdvancedTransportPhenomena2007}:
\begin{subequations}
\begin{align}
    \boldsymbol{v}^\mathrm{HI}(\boldsymbol{r})  &= \boldsymbol{u}^{*}_{\mathrm{FD}}(\boldsymbol{r}) = \frac{3\alpha}{16z^2}\left(1-3\cos(2\vartheta)\right)\boldsymbol{\hat{z}}+\frac{3\alpha}{8z^2}\sin(2\vartheta)\boldsymbol{\hat{\phi}},  \\
     \quad \boldsymbol{\omega}^\mathrm{HI} (\boldsymbol{r}) &= \frac{1}{2}\boldsymbol{\nabla}\wedge \boldsymbol{u}^{*}_{\mathrm{FD}}(\boldsymbol{r}) = \frac{3\alpha}{16z^3}\sin(2\vartheta)\boldsymbol{\hat{\phi}}^{\perp},
\end{align}
\end{subequations}
with $\boldsymbol{\hat{\phi}}^{\perp} = \boldsymbol{\hat{z}}\wedge\boldsymbol{\hat{\phi}}$.

\subsection{Drift \label{app:drift}}
We follow Ref.~\cite{jonesRotationalDiffusionTracer1992a} to derive the equations of motion for the position $\boldsymbol{r}$ and orientation $\boldsymbol{e}$ of the active particle. Therefore, we consider a particle with spatially-dependent translational and rotational diffusivities and ignore any deterministic terms (i.e., active motion and hydrodynamic reorientation and attraction) at this point, as the latter can be simply added. The associated Fokker-Planck equation for the probability density $p(\boldsymbol{x},t|\boldsymbol{x}_0)$, which depends on the particle's generalized instantaneous configuration $\boldsymbol{x}=(\boldsymbol{r},\boldsymbol{e})$ and initial configuration $\boldsymbol{x}_0=(\boldsymbol{r}_0,\boldsymbol{e}_0)$, reads: 
\begin{align} \label{eq:fp}
\partial_t p = \boldsymbol{\nabla}\cdot\left(\boldsymbol{D}\cdot\boldsymbol{\nabla}p\right),
\end{align}
where $\boldsymbol{D}$ denotes the generalized diffusion tensor 
\begin{align}
\boldsymbol{D} &= \begin{pmatrix} \boldsymbol{D}^{t} & \boldsymbol{D}^{tr}\\
\boldsymbol{D}^{rt}  & \boldsymbol{D}^{r}
\end{pmatrix} \qquad \mathrm{with} \qquad  \boldsymbol{D}^{rt}  = (\boldsymbol{D}^{tr})^T.
\end{align}
The latter depends on the spatial coordinate $\boldsymbol{D}= \boldsymbol{D}(\boldsymbol{r})$ only [Eqs.~(5a)-(5d) in the main text]. Furthermore, the generalized gradient assumes the form $\boldsymbol{\nabla}= (\partial_{\boldsymbol{r}}, \boldsymbol{e}\wedge\partial_{\boldsymbol{e}})$, accounting for the fact that the orientation is normalized $|\boldsymbol{e}|=1$. We next consider the initial condition $p(\boldsymbol{x},0|\boldsymbol{x}_0)=\delta(\boldsymbol{x}-\boldsymbol{x}_0)$ and can thus formally write the solution as
\begin{align}
p(\boldsymbol{x},t|\boldsymbol{x}_0) = e^{\mathcal{D}t}\delta(\boldsymbol{x}-\boldsymbol{x}_0) 
\end{align}
with operator $\mathcal{D} = \boldsymbol{\nabla}\cdot\boldsymbol{D}\cdot\boldsymbol{\nabla}$. For infinitesimally small times $\Delta t$, this expression can be expanded as
\begin{align}
p(\boldsymbol{x},t|\boldsymbol{x}_0) = \delta(\boldsymbol{x}-\boldsymbol{x}_0)+\Delta t  \mathcal{D} \delta(\boldsymbol{x}-\boldsymbol{x}_0) +O\left(\Delta t^2\right),
\end{align}
allowing to compute averages of observables $A(t)$ to order $\Delta t$ as
\begin{subequations}
\begin{align}
\langle A(\boldsymbol{x})\rangle &= A(\boldsymbol{x}_0) + \Delta t \int A(\boldsymbol{x})\mathcal{D}\delta(\boldsymbol{x}-\boldsymbol{x}_0) \mathrm{d}\boldsymbol{x}   +O\left(\Delta t^2\right), \\
&= A(\boldsymbol{x}_0) + \Delta t \mathcal{D}^\dagger A(\boldsymbol{x})|_{\boldsymbol{x}=\boldsymbol{x}_0} +O\left(\Delta t^2\right), 
\end{align}
\end{subequations}
with adjoint operator $\mathcal{D}^\dagger$ ($=\mathcal{D}$ in the absence of a drift in Eq.~\eqref{eq:fp}). We note that 
\begin{subequations}
\begin{align}
\mathcal{D} A(\boldsymbol{r}, \boldsymbol{e}) &= (\partial_{\boldsymbol{r}}, \boldsymbol{e}\wedge\partial_{\boldsymbol{e}}) \cdot \left[\begin{pmatrix}
\boldsymbol{D}^{t} & \boldsymbol{D}^{tr}\\
\boldsymbol{D}^{rt}  & \boldsymbol{D}^{r}
\end{pmatrix} \cdot \left[(\partial_{\boldsymbol{r}}, \boldsymbol{e}\wedge \partial_{\boldsymbol{e}}) A(\boldsymbol{r}, \boldsymbol{e}))\right]\right],\\
&= \partial_{\boldsymbol{r}}\cdot \left(\boldsymbol{D}^{t}\cdot \partial_{\boldsymbol r} A +\boldsymbol{D}^{tr} \cdot(\boldsymbol{e}\wedge \partial_{\boldsymbol{e}}A)\right) + \boldsymbol{e}\wedge \partial_{\boldsymbol{e}}\cdot \left(\boldsymbol{D}^{rt}\cdot\partial_{\boldsymbol{r}}A+\boldsymbol{D}^{r}\cdot(\boldsymbol{e}\wedge \partial_{\boldsymbol{e}} A)\right).
\end{align}
\end{subequations}
and compute the leading-order instantaneous translational and angular velocities via 
\begin{subequations}
\begin{align} 
\lim_{\Delta t \to 0} \frac{\langle \boldsymbol{r}-\boldsymbol{r}_0\rangle}{\Delta t} &= \partial_{\boldsymbol{r}}\cdot \boldsymbol{D}^{t}\cdot \partial_{\boldsymbol r} \boldsymbol{r}|_{\boldsymbol{x}=\boldsymbol{x}_0}  = \partial_{\boldsymbol{r}}\cdot \boldsymbol{D}^{t}|_{\boldsymbol{x}=\boldsymbol{x}_0} \label{eq:dr}\\
\lim_{\Delta t \to 0} \frac{\langle \boldsymbol{e}-\boldsymbol{e}_0\rangle}{\Delta t} &= \boldsymbol{e}\wedge \partial_{\boldsymbol{e}}\cdot \left(\boldsymbol{D}^{r}\cdot \left(\boldsymbol{e}\wedge \partial_{\boldsymbol e} \boldsymbol{e}\right)\right)|_{\boldsymbol{x}=\boldsymbol{x}_0} = \boldsymbol{D}^{r}\cdot \boldsymbol{e}_0-\mathrm{Tr}[\boldsymbol{D}^{r}]\cdot\boldsymbol{e}_0 + \partial_{\boldsymbol{r}} \cdot \boldsymbol{D}^{rt} \wedge \boldsymbol{e}_0, \label{eq:de}
\end{align}
\end{subequations}
where $\mathrm{Tr}$ denotes the trace. As $\boldsymbol{D}$ does not depend on $\boldsymbol{e}$, we find that the translational/rotational coupling  does not enter Eq.~\eqref{eq:dr}. In addition, the terms arising in  Eq.~\eqref{eq:de} do not arise from the spatial dependence of the diffusivities but because reorientation due to rotational diffusion is anisotropic. This feature has already been pointed out for the rotational diffusion of an anisotropic, active particle~\cite{kurzthalerIntermediateScatteringFunction2016}, while it arises here for a spherical particle due to the presence of a wall. In our work we ignore the translational/rotational coupling ($\boldsymbol{D}^{rt}=\boldsymbol{0}$) and further express the result in terms of components parallel and perpendicular to $\boldsymbol{e}_0$
\begin{align}
   \lim_{\Delta t \to 0} \frac{\langle \boldsymbol{e}-\boldsymbol{e}_0\rangle}{\Delta t} = \Big( \boldsymbol{e}_0 \cdot \boldsymbol{D}^{r} \cdot \boldsymbol{e}_0 - \mathrm{Tr}[\boldsymbol{D}^{r}] \Big) \boldsymbol{e}_0 + \Big( \boldsymbol{e}_0 \wedge \boldsymbol{D}^{r} \cdot \boldsymbol{e}_0 \Big) \wedge \boldsymbol{e}_0 \label{eq:de2}
\end{align}
where the first term enforces normalization of the orientation $\boldsymbol{e}$ and can be discarded by normalizing the orientation at each simulation time step while the second term induces a reorientation  arising from the anisotropy of the diffusion tensor $\boldsymbol{D}^{r}$~\cite{jonesRotationalDiffusionTracer1992a}. Decomposing the diffusion matrix as
\begin{align}
    \boldsymbol{D}^{r} = D_{\perp}^{r} \boldsymbol{\hat{z}} \boldsymbol{\hat{z}} + D_{\parallel}^{r} (\boldsymbol{I} - \boldsymbol{\hat{z}}\boldsymbol{\hat{z}}),
\end{align}
we note that the geometric drift can be further simplified to
\begin{align}
   \Big( \boldsymbol{e} \wedge \boldsymbol{D}^{r} \cdot \boldsymbol{e} \Big) \wedge \boldsymbol{e} =  -\Big( D_{\perp}^{r} - D_{\parallel}^{r} \Big) \left(e_{z}^{2} \boldsymbol{e} - e_{z}\boldsymbol{\hat{z}} \right).
\end{align}
The mean velocities [Eqs.~\eqref{eq:dr},\eqref{eq:de2}] represent spurious drift terms (due to both spatially varying diffusivities and anisotropic orientational diffusion) in the associated Langevin equations (in the It\=o sense) and assure that the distribution assumes the equilibrium Boltzmann distribution.

\subsection{Non-dimensionalization of the equations of motion}

Rescaling the equations of motion with the particle radius $a$ as characteristic length scale, $\boldsymbol{r}=a\boldsymbol{R}$, and the rotational diffusion time $\tau = 1/(2D_0^{r})$ as characteristic time scale, $t = \tau T$, yields 

\begin{subequations}
\begin{align}
    \frac{\mathrm{d} \boldsymbol{R }}{\mathrm{d} T} &= \mathrm{Pe}\boldsymbol{e} + \mathrm{Pe} \hat{\alpha}\boldsymbol{V}^\mathrm{HI} + \boldsymbol{V}^\mathrm{noise}+ \boldsymbol{V}^\mathrm{drift}, \label{eq:dynamic_position_nodim} \\
    \frac{\mathrm{d} \boldsymbol{e}}{\mathrm{d} T} &= \left(\mathrm{Pe} \hat{\alpha}\boldsymbol{\Omega}^\mathrm{HI}+\boldsymbol{\Omega}^\mathrm{noise} +\boldsymbol{\Omega}^{\mathrm{drift}}\right)\wedge\boldsymbol{e} , \label{eq:dynamic_angle_nodim}
\end{align}
\end{subequations}
with 
\begin{subequations}
\begin{align}
\boldsymbol{V}^{\mathrm{HI}} &=\frac{3}{8Z^{2}}\sin(2\vartheta) \left( \cos(\varphi_{0})\boldsymbol{\hat{x}} + \sin(\varphi_{0})\boldsymbol{\hat{y}}\right) + \frac{3}{16Z^2}\left(1-3\cos(2\vartheta)\right)\boldsymbol{\hat{z}}\\ 
\boldsymbol{\Omega}^{\mathrm{HI}} &= \frac{3}{16Z^{3}}\sin(2\vartheta) \left( -\sin(\varphi_{0}) \boldsymbol{\hat{x}} + \cos(\varphi_{0})\boldsymbol{\hat{y}} \right), \\ 
\boldsymbol{V}^{\mathrm{drift}} &= \frac{3}{4Z^{2}} \hat{\boldsymbol{z}}, \\
\boldsymbol{\Omega}^{\mathrm{drift}} \wedge \boldsymbol{e} &= -  (\mathcal{D}_{\perp}^r(Z)-\mathcal{D}_{\parallel}^r(Z))({e_z}^2 \boldsymbol{e} - {e_z}\boldsymbol{\hat{z}}).
\end{align}
\end{subequations}
and rescaled diffusivities
\begin{subequations}
\begin{align}
    \mathcal{D}_{\perp}^{t}(Z) &= \frac{2}{3}\left(1 -  \frac{9}{8Z}\right), &\quad
    \mathcal{D}_{\parallel}^{t}(Z) &= \frac{2}{3}\left(1 -  \frac{9}{16Z}\right),  \\
    \mathcal{D}_{\perp}^{r}(Z) &= \frac{1}{2}\left (1 - \frac{1}{8Z^{3}}  \right ), &\quad 
    \mathcal{D}_{\parallel}^{r}(Z) &= \frac{1}{2} \left (1 - \frac{5}{16Z^{3}} \right ),
\end{align}
\end{subequations}
where we used the Stokes-Einstein-Sutherland relation for a spherical particle in bulk $D_{0}^{t}/D_{0}^{r} = (4/3)a^{2}$.

\subsection{Numerical aspects}

The equations of motion are integrated with the following scheme:
\begin{subequations} 
    \begin{align}
        \mathbf{R}(T+\Delta T) &= \mathbf{R}(T) + \Delta T \left(\mathrm{Pe} \boldsymbol{e}(T) + \mathrm{Pe}\hat{\alpha} \boldsymbol{V}^{\mathrm{HI}}  + \boldsymbol{V}^{\mathrm{drift}}\right) + \mathrm{diag}\left (\sqrt{2\mathcal{D}_{\parallel}^{t}(Z)\Delta T}, \sqrt{2\mathcal{D}_{\parallel}^{t}(Z)\Delta T}, \sqrt{2\mathcal{D}_{\perp}^{t}(Z)\Delta T} \right) \cdot  \boldsymbol{N}(0, 1), \\ 
        \boldsymbol{e}(T+\Delta T) &= \boldsymbol{e}(T) - \Delta T \left( \mathcal{D}_{\perp}^{r}(Z)  - \mathcal{D}_{\parallel}^{r}(Z)  \right)\left({e_z}^2 \boldsymbol{e}(T) - {e_z}\boldsymbol{\hat{z}} \right)  \notag\\
        & \qquad +\left( \mathrm{Pe} \hat{\alpha
        } \boldsymbol{\Omega}^{\mathrm{HI}}(T) \Delta T + \mathrm{diag} \left (\sqrt{2\mathcal{D}_{\parallel}^{r}(Z)\Delta T}, \sqrt{2\mathcal{D}_{\parallel}^{r}(Z)\Delta T}, \sqrt{2\mathcal{D}_{\perp}^{r}(Z)\Delta T} \right) \cdot  \boldsymbol{\xi}(0, 1)  \right) \wedge \boldsymbol{e}(T), \\ 
        \boldsymbol{e}(T+\Delta T) &= \frac{\boldsymbol{e}(T+\Delta T)}{\lvert \boldsymbol{e}(T+\Delta T) \rvert},
    \end{align}
\end{subequations}
where $\boldsymbol{\xi}(0, 1)$ and $\boldsymbol{N}(0, 1)$ are independent, normally distributed random variable with zero mean and unit variance. Furthermore we use a time-step $\Delta T = 10^{-4}$ and collect statistics from $10^{5}$ agents.

\section{Additional results}

\subsection{Active Brownian particle}

For reference, we show the median first-passage time (FPT) $T_{1/2}$ of an active Brownian particle (corresponding to the absence of hydrodynamic effects) as a function of the initial distance $z_{0}$. Its dynamics corresponds to the case $\boldsymbol{V}^{\mathrm{HI}}=\boldsymbol{\Omega}^{\mathrm{HI}}= \boldsymbol{0}$ with constant bulk translational and rotational diffusivities $D_{0}^{t}$ and $D_{0}^{r}$, respectively, so that the spurious drift term vanishes. At very short $z_{0}/a \ll 1$ and very large $z_{0}/a \gg1 $ initial distances, the median scales as $z_{0}^{2}$, similar to a Brownian particle, owing to the initial diffusive regime (with diffusivity $D_{0}^{t}$) and effective diffusive regime at long times (with effective diffusion coefficient $D_{\mathrm{eff}}= D_{0}^{t} + v^{2}/(6D_{0}^{r})$). For intermediate initial distances, the median deviates from the quadratic scaling and behaves suballistically $T_{1/2} \propto z_{0}^{2/3}$, as explained in the main text.

\begin{figure}[htp]
\includegraphics[scale=0.72]
{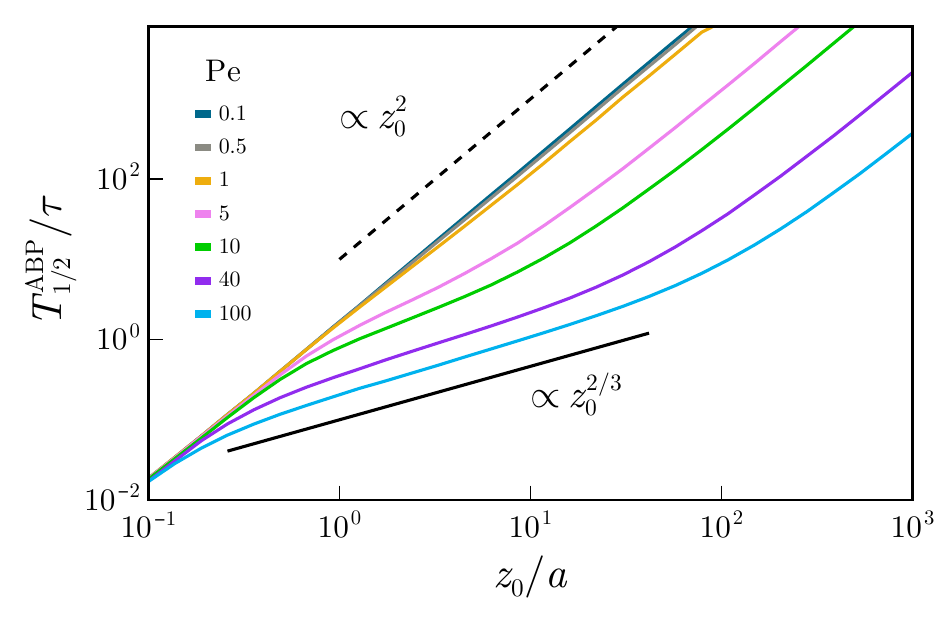}
\caption{Median FPT $T_{1/2}$ of a ``dry'' active Brownian particle as a function of the initial distance $z_0/a$ for various P{\'e}clet numbers. \label{fig:median_abp}}
\end{figure}

\subsection{Split role of diffusivities}

\begin{figure}[h!]
\includegraphics[width =0.9\textwidth]
{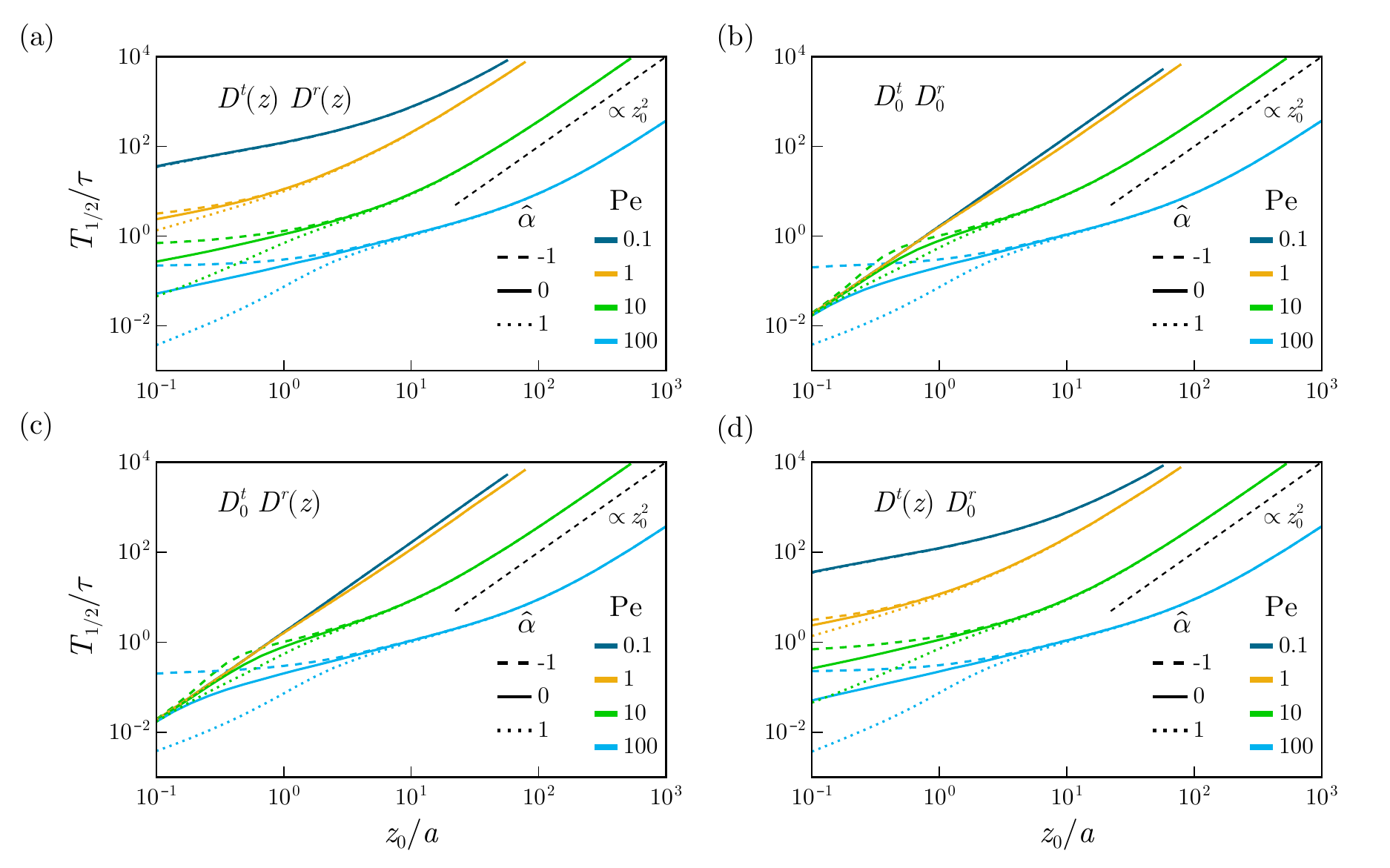}
\caption{Median FPT $T_{1/2}$ as a function of the initial distance $z_{0}$ for several P{\'e}clet numbers $\mathrm{Pe}$ and dipole strengths $\hat{\alpha}$. (a) Both translational and orientational diffusion coefficients are $z$-dependent. (b) Both translational and orientational diffusion coefficients are constant and equal the one in the bulk. (c) The rotational diffusion coefficient is $z$-dependent while the translational diffusion coefficient is the bulk one. (d) The translational diffusion coefficient is $z$-dependent while the rotational diffusion coefficient is the bulk one.\label{fig:role_diffusivities}}
\end{figure}

Here, we show the median FPT for one or several constant diffusivities. Figure~\ref{fig:role_diffusivities}(a) is the reference case with both translation and rotational diffusivities that are $z$-dependent. Taking a constant rotational diffusivity [Fig.~\ref{fig:role_diffusivities}(d)] presents no difference with the reference case, thus showing that spatially-varying rotational diffusion plays a negligible role in the FPT statistics (Note that it enters the equations of motion at a higher order in $a/z$.). In addition, only keeping $D_{0}^{t}$ or $D_{0}^{t}$ and $D_{0}^{r}$ makes a big difference, as one retrieves the ABP behavior for neutral swimmers, while hydrodynamic interactions still play their role at high activity.

\subsection{Impact of initial angle}

\begin{figure}[h!]
\includegraphics[width =0.8\textwidth]
{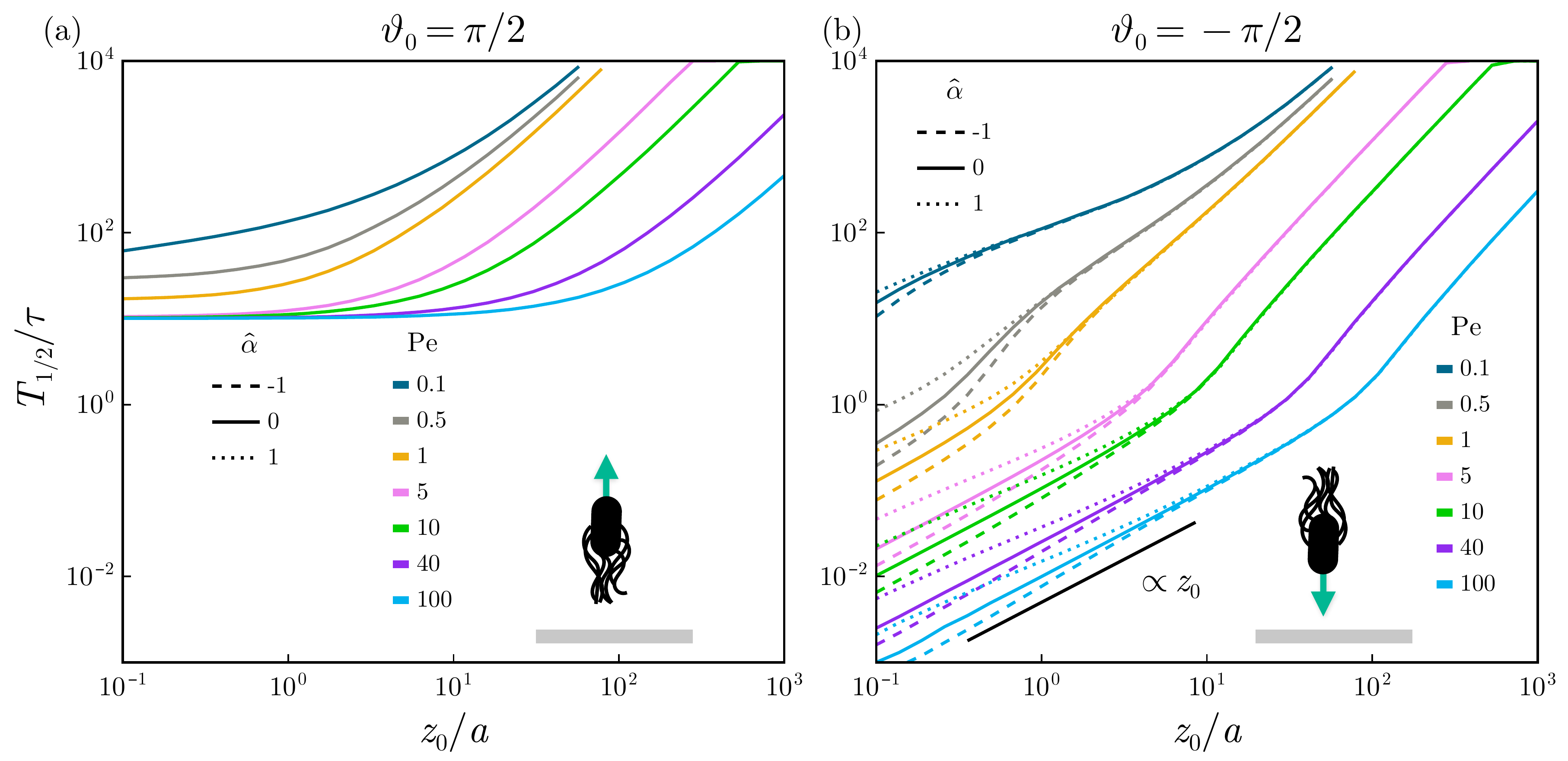}
\caption{Median FPT $T_{1/2}$ as a function of the initial distance $z_{0}$ for several P{\'e}clet numbers and dipole strengths $\hat{\alpha}$. The swimmer is initially oriented (a)~away from the wall $(\vartheta_{0}=\pi/2)$ and (b)~towards the wall $(\vartheta_{0}=-\pi/2)$.  \label{fig:median_angles}}
\end{figure}

Here, we plot the median FPT $T_{1/2}$ as function of the initial distance $z_{0}$ for the case where the agent initially faces away from the boundary [Fig.~\ref{fig:median_angles}(a)] and where it initially faces the boundary  [Fig.~\ref{fig:median_angles}(b)]. The first case illustrates how activity reduces the median FPT up to a certain point ($\mathrm{Pe}\simeq 5$), where $T_{1/2}$ plateaus to a certain value. Highly active swimmer can only reach the boundary by reorientating through orientational diffusion, therefore this plateau lasts longer as $\mathrm{Pe}$ (and thus the persistence length $l_{p}$) increases.
The case where agents start to initially swim towards the boundary shows that the median initially behaves ballistically for neutral swimmers, owing to the directed persistent motion. It also exposes how hydrodynamic interactions attract pullers and repel pushers at short distances, as the sign of the $z$ component of $( \mathrm{Pe}\hat{\alpha}\boldsymbol{V}^{\mathrm{HI}})$ is dictated by $\hat{\alpha}$ for $\vartheta= -\pi/2$.

\subsection{Distribution of arrival angles}
\begin{figure}[h!]
\includegraphics[width =\textwidth]
{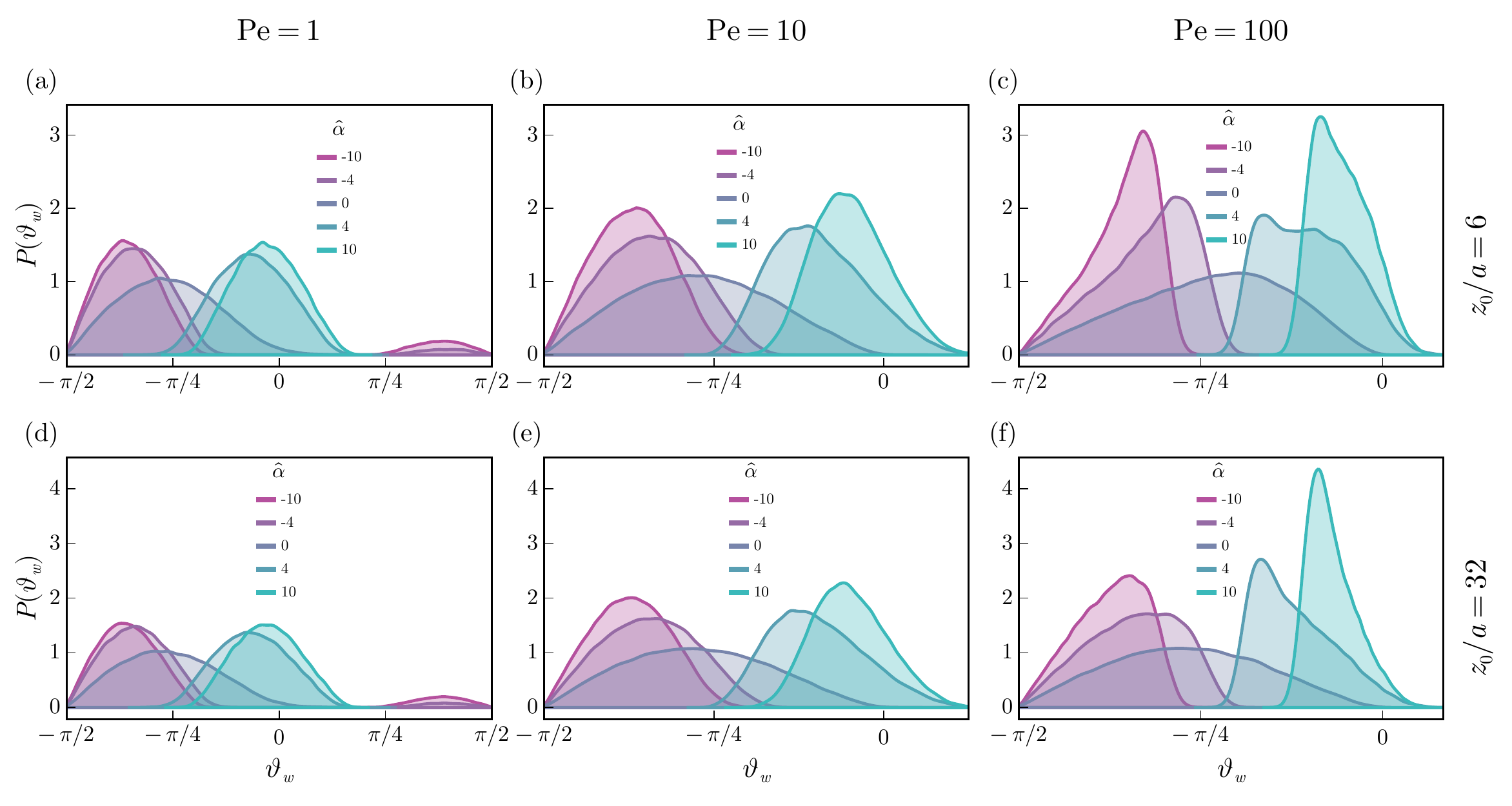}
\caption{\label{fig:angle_distrib} Distribution of the arrival angles $\vartheta_{w}$ as a function of the dipole strength $\hat{\alpha}$ for several P{\'e}clet numbers and initial distances $z_{0}$. First row (a)-(b)-(c) is for an initial distance $z_{0}/a=6$ and the second row (d)-(e)-(f) for $z_{0}/a=30$.}
\end{figure}

Here, we show how the distribution of arrival angles varies with activity ($\mathrm{Pe}$) and initial distance $z_{0}/a$ [Fig.~\ref{fig:angle_distrib}]. For low activity ($\mathrm{Pe}=1$) [Fig.~\ref{fig:angle_distrib}(a),(d)],  the swimmer can reach the wall while being aligned opposed to it, as the dynamics is heavily influenced by diffusion. Increasing the P{\'e}clet number  or the initial position comforts the prediction that pushers tend to arrive parallel to the boundary (shift towards $\vartheta_{w} =0 $) while pullers tend to arrive perpendicular (shift towards $\vartheta_{w} = -\pi/2$) [Fig.~\ref{fig:angle_distrib}(b-c)-(e-f)].

\subsection{Split role of hydrodynamic contributions}
\begin{figure}[h!]
\includegraphics[width =0.8\textwidth]
{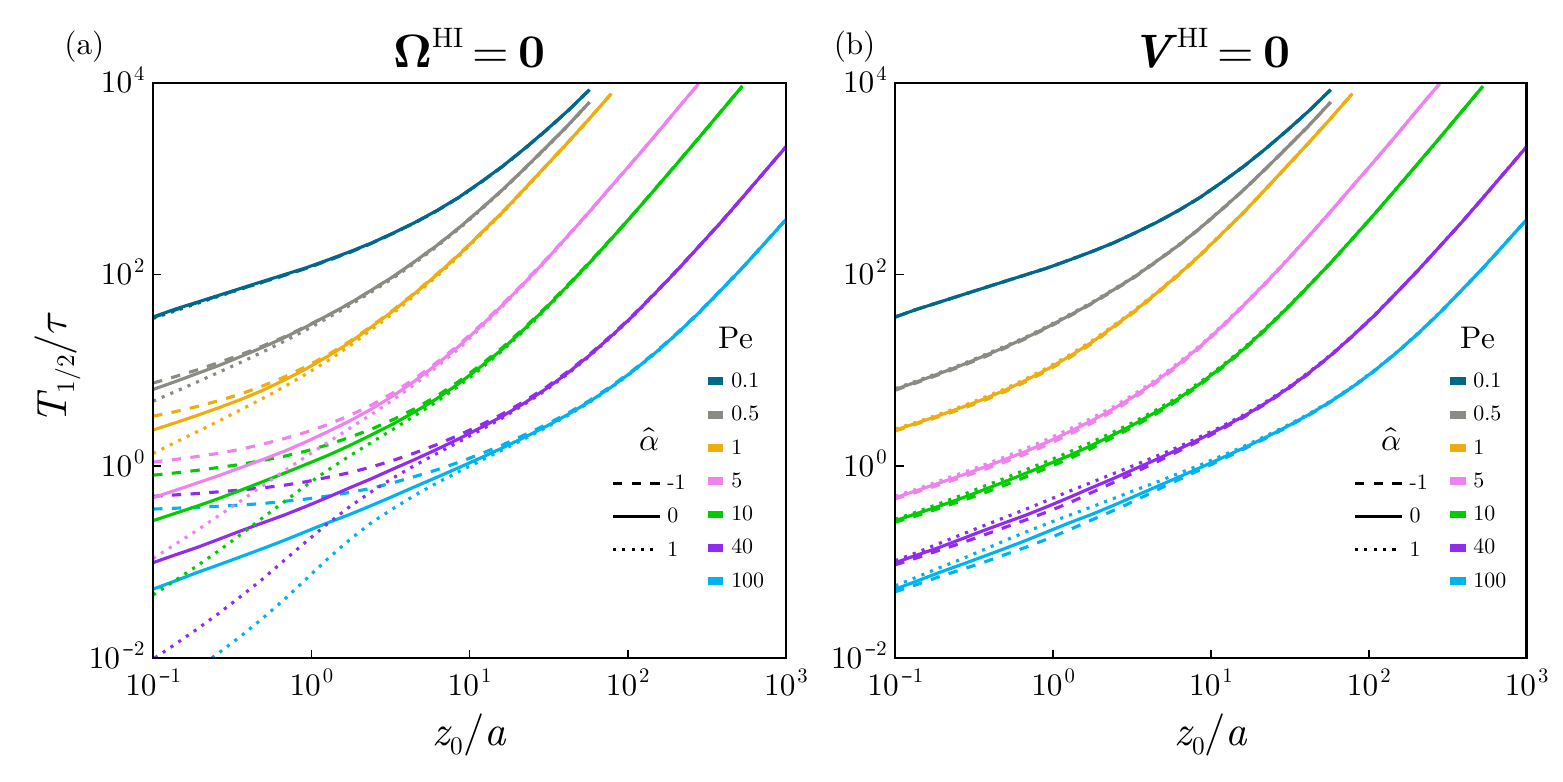}
\caption{\label{fig:hydro_contrib} Median FPT $T_{1/2}$ as a function of the initial distance $z_{0}$ for several P{\'e}clet numbers and dipole strengths $\hat{\alpha}$. (a) The hydrodynamic-induced torques are switched off $\boldsymbol{\Omega}^{\mathrm{HI}} =\boldsymbol{0}$. (b) The hydrodynamic-induced velocities are switched off $\boldsymbol{V}^{\mathrm{HI}} =\boldsymbol{0}$. }
\end{figure}
Finally, we show how the individual hydrodynamic contributions affect the median. Not considering the hydrodynamic torques [$\boldsymbol{\Omega}^{\mathrm{HI}}=0$ in Fig.~\ref{fig:hydro_contrib} (a)] does not yield visible changes, thus indicating that the hydrodynamic attraction and repulsion are the main factors of discrepancy between pushers and pullers. On the other hand, switching off the speed contribution [($\boldsymbol{V}^{\mathrm{HI}}=0$) in Fig.~\ref{fig:hydro_contrib} (b)] makes the pullers slightly faster, as they tend to reorient perpendicular to the boundary.

\bibliography{bibliography_supplementary}